\documentstyle[twoside,psfig]{adacta}

\def\autor{S. Stenholm and K.-A. Suominen}
\def\nazov{The wave packet - a universal quantum object}
\begin{document}

\headings{1}{8}

\title{THE WAVE PACKET - A UNIVERSAL QUANTUM OBJECT}

\author{S. Stenholm}{The Academy of Finland and
Helsinki Institute of Physics, PL 9, 00014 Helsingin yliopisto, Finland}

\author{K.-A. Suominen}{Theoretical Physics
Division, Department of Physics, University of Helsinki, PL 9, 00014 Helsingin
yliopisto, Finland}

\abstract{We summarize the theoretical description of wave packets on molecular
energy levels. We review the various quantum mechanical effects which can be
studied and the models that can be verified on this system. This justifies
our claim that the wave packet constitutes a universal quantum object.}

\section{Introduction}

Quantum Mechanics has turned out to be our most versatile and universal
description of microscopic Nature. Its concepts and methods have been applied
to ever larger ranges of microscopic entities and their accompanying
phenomena. However, most applications are based on calculating the stationary
energy eigenvalues or the probability fluxes in scattering situations. The
theory predicts a multitude of energy levels in atoms, molecules, solids,
nuclei and elementary particles. Also, the standard approach to scattering
utilizes a steady flow of particles instead of the normalizable wave packets
corresponding more closely to the physical conditions.

Recently the situation has changed considerably. The technical development
of well controlled laser pulses has made it possible to excite
non-stationary quantum states and follow their Schr\"{o}dinger evolution in
real time. Similar experiments were earlier carried out with nuclear and
electronic spins, but their time evolution takes place in a finite Hilbert
space, whereas only experiments on atoms and molecules can address  excited
states living on a continuum. 

With picosecond laser pulses one can excite wave packets on Rydberg levels
and probe their quantum fate, see~\cite{bx1} and for a review~\cite{bx2}.
With the recent development of femtosecond pulses, one can excite nuclear wave
packets on the adiabatic electronic levels of molecules~\cite{ba1} and follow
their motion. Within the Born-Oppenheimer description, this is the ideal
laboratory for wave packet experiments~\cite{bb1}, and it provides an excellent
opportunity to test the validity of much used models in quantum theory; see
Ref.~\cite{b1} where much of our work on wave packets is reviewed. We have also
applied time dependent methods to electronic states in semiconductor
heterostructures~\cite{ba2}, where the potential wells provide an interesting
analogy to the energy levels of molecules. Furthermore, these methods have been
very useful in studies of cold collisions between laser cooled and trapped
neutral atoms~\cite{JPB-KAS}.

In this paper, we will review the physical formulation of laser-induced
processes in molecules (Sec.~\ref{formulate}), summarize some of the physical
phenomena investigated (Sec.~\ref{physics}), and present some of our most
recent results (Sec.~\ref{decay}). These are mainly found in the work~\cite{bc1}
being the PhD Thesis of Asta Paloviita.

\section{Formulation of the problem}\label{formulate}

Within the Born-Oppenheimer approximation, we solve for the electronic
energy levels with the nuclei localized at their classical positions $R$.
The corresponding eigenvalues ${\cal U}_n(R)$ give the potential surfaces
for the nuclear motion, and the corresponding eigenfunctions $\varphi_n(r)$
form an orthonormal set in the space of electronic states. This can be
utilized as a basis even when the electronic levels are coupled by external
laser fields. We thus write for the full state of the system the wave
function
\begin{equation}
  \Psi (r,R;t)=\sum_n\psi _n(R,t)\,\varphi _n(r),  \label{e1}
\end{equation}
which is still exact. In the Born-Oppenheimer approximation, we neglect the 
$R$-dependence of the electronic states $\varphi_n(r)$, and obtain the
equations 
\begin{equation}
   \begin{array}{lll}
   i\hbar \frac \partial {\partial t}\psi _n(R,t) & = & 
   \left[ -\frac{\hbar^2}{2M}\nabla ^2+{\cal U}_n(R)\right] \psi _n(R,t) \\ 
   &  &  \\ 
   &  & +\sum_m\langle \varphi _n\mid {\bf D\mid }\varphi _m\rangle \cdot 
   {\bf E}(t)\;\psi _m(R,t).
   \end{array}\label{e2}
\end{equation}
Here ${\bf E}(t)$ is the external field coupled to the molecular operator
{\bf D}. Near resonance between two molecular energy levels, we can perform
the rotating-wave approximation and reduce the problem to the coupled
equations
\begin{equation}
   \begin{array}{lll}
   i\frac \partial {\partial t}\psi _1 & = & \left[ -\frac{\partial ^2}{
   \partial x^2}+{\cal U}_1(x)\right] \psi _1+V\,\psi _2 \\ 
   &  &  \\ 
   i\frac \partial {\partial t}\psi _2 & = & \left[ -\frac{\partial ^2}{
   \partial x^2}+{\cal U}_2(x)-\omega \right] \psi _2+V\,\psi _1.
   \end{array} \label{e3}
\end{equation}
Here $\omega$ is the laser frequency, and we have scaled the variables in a
way adapted to numerical computations. Most of our work has been based on
equations like~(\ref{e3}); to introduce one more level is simple, but to add
more dimensions strains even up-to-date computer resources. For a discussion
of the numerical problems encountered see Ref.~\cite{b1}.

Solving the coupled equations~(\ref{e3}) directly does not involve the
Franck-Condon principle, which, however, is embedded in the local nature of
the excitation process. Only for distances satisfying an energy resonance
condition, can population transfer between the levels take place. The Condon
factors can be made explicitly visible by using an expansion in the
nuclear eigenfunctions on the electronic energy surfaces, see e.g.~\cite{by1}.

A straightforward application of the equations~(\ref{e3}) is an investigation
of the level crossing models introduced by Landau and Zener. By tuning the
laser frequency $\omega$, we can move the crossing point given by
\begin{equation}
   {\cal U}_2(x)-\omega ={\cal U}_1(x),  \label{e4}
\end{equation}
and the coupling strength $V$ is determined by the laser intensity. In this
situation we have been able to verify the Landau-Zener theory by integrating
the wave packet motion across a laser-induced crossing \cite{b2}. In fact,
our computations show that it is difficult to find experimentally
significant deviations from the simple theory.

\section{Physics of wave packet excitation}\label{physics}

The ground state of a molecule prepares nearly an exact Gaussian wave
packet, which can be lifted to an excited level by a laser pulse, see
Fig.~1. Ideally this can take place without
distortion, but quantum dynamics distorts the shape for pulses of finite
duration~\cite{b3}. 

\begin{figure}[htb]
\centerline{\psfig{width=80mm,file=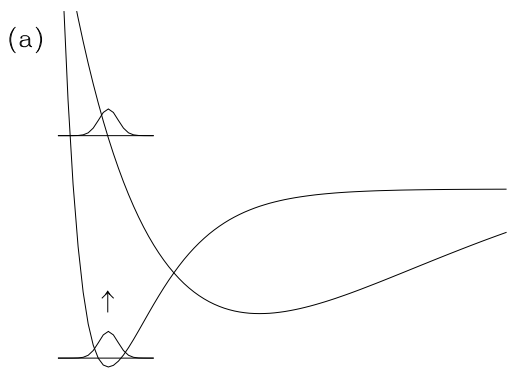}}
\caption{Fig. 1: The basic model for molecular excitation by
ultrashort pulses.\label{asta}}
\end{figure}

At once upon excitation, the excited wave packet starts to spread, in a manner
determined mainly by the initial width of the wave packet. The wings of
a broad wave packet are out of resonance, and when the population starts to
return back to the ground state, the wings do not participate in the Rabi
flopping occurring during the laser pulse. If the wave packet is excited to
a potential slope, the situation is less straightforward, and hence
the emerging wave packet may assume a rather complicated shape as shown
in~\cite{b3}. Near the resonance position, we may see the standard Rabi
flopping between the two levels. 

If the slope on the excited level is steep enough, the excited wave packet is
rapidly accelerated and thus it escapes the resonance region. Then it can no
longer participate in the flopping and is returned less efficiently to the
ground state. By chirping the laser frequency in a propitious way, one may
increase the excitation efficiency of the process~\cite{by2}.

When the intensity and the length of the laser pulse are increased, the
population on the upper level has not enough time to undergo any quantum
dynamics before it is returned to the ground level. Thus on the upper state,
the motion appears frozen~\cite{b5}, and only the fraction of the population
remaining at the end of the pulse can escape as shown in Fig.~2.

Another phenomenon disrupting the excitation process is spontaneous decay.
When the decay time is of the same order as the exciting pulse, this allows
some excitation followed by eventual return to the ground state. We have
investigated the dynamics of this situation~\cite{b4} using the state vector
Monte Carlo method. This also gives the statistics of the position where the
spontaneous decay occurs; as this determines the central frequency of the
emerging photon, we are thus able to provide a picture of the time dependent
spectrum too~\cite{bb4}.
\nopagebreak
\begin{figure}[htb]
\vspace*{-10cm}
\centerline{\psfig{width=120mm,angle=-90,file=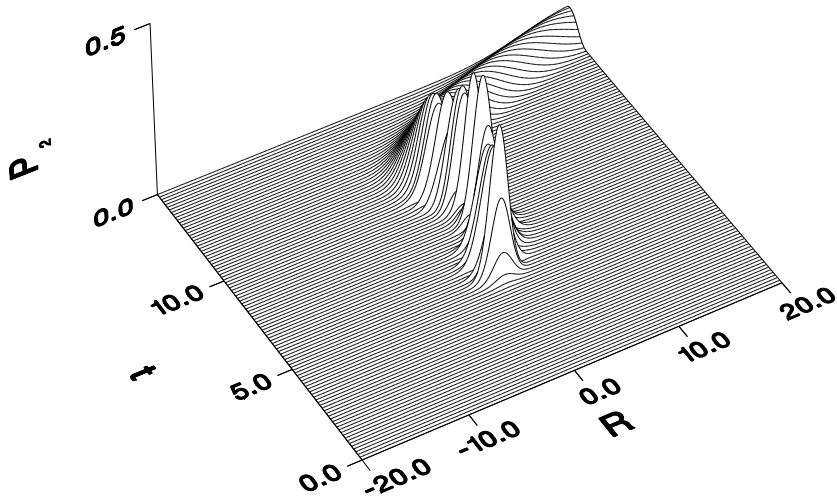}}
\vspace*{7cm}
\caption{Fig. 2: The excited state wave packet $P_2(R,t)$ during an excitation
by a pulse with a Gaussian envelope.\label{d3gauss}}
\end{figure}

\section{Wave packet decay}\label{decay}

When the ground state of a molecule is coupled suddenly to the slope of an
excited state, its discrete bound energy level becomes coupled to the
continuum of the sloping potential. This is exactly the situation treated by
Weisskopf and Wigner~\cite{b6} as a model of exponential decay. Such a
model has become the prototype for radiative decay, particle creation and
resonances in scattering theory \cite{b7}. It is also used for describing
non-radiative processes in large molecules \cite{b8} and dissipative
processes in Quantum Optics \cite{bb8}. 

The simple case of a discrete state $|0\rangle$ embedded in a
continuum $\{|\epsilon \rangle \}$ and coupled linearly to it, can be
described by the model
\begin{equation}
   \begin{array}{lll}
   H & = & H_0+V \\ 
   &  &  \\ 
   H_0 & = & \omega _0| 0\rangle \langle 0| +\int d\epsilon | \epsilon
   \rangle \,\epsilon \,\langle \epsilon | \\ 
   &  &  \\ 
   V & = & \int d\epsilon \left( V_\epsilon | \epsilon \rangle \langle 0\mid
   +V_\epsilon ^{*}| 0\rangle \langle \epsilon | \right) .
   \end{array}\label{f1}
\end{equation}
This system can be solved using the standard methods of quantum
theory~\cite{b9},  and with the initial state $\varphi _0(x)\equiv \langle x|
0\rangle$ we find the time evolution of this state to be
\begin{equation}
   \psi (x,t)=\exp \left[ -\left( i\omega'+\Gamma /2\right)
   t\right] \varphi _0(x),  \label{f2}
\end{equation}
where $\omega'$ is the renormalized frequency $\omega _0$ and $
\Gamma $ is the decay rate; in the weak coupling limit this is given by the
Weisskopf-Wigner expression. The wave packet on the continuum levels is
found to emerge in the form
\begin{equation}
   \Psi (x,t)=\Phi (x,t)-\exp \left[ -\left( i\omega'+\Gamma
   /2\right) t\right] \Phi (x,0),  \label{ff2}
\end{equation}
where $\Phi (x,t)$ is the outgoing wave packet for times much longer than $
\Gamma^{-1}$. The validity of this result and its correction terms are
discussed in Ref.~\cite{b9}.

To test the validity of the expression~(\ref{f2}), we have replaced the
simple model~(\ref{f1}) by one of the type~(\ref{e3}) with the potentials
\begin{equation}
   \begin{array}{lll}
   {\cal U}_1(x) & = & \frac{1}{2}x^2 \\ 
   &  &  \\ 
   {\cal U}_2(x)-\omega  & = & \frac{1}{\sqrt{2}}-\alpha x.
   \end{array}\label{f3}
\end{equation}
With this choice, we find the energies of the discrete state to coincide
with that of the continuum at $x=0$, see Fig.~3. 
When we couple the two, we find that the decayed population emerges on the
slope as a localized wave packet, Fig.~4. The
probability of remaining on the original level is shown in Fig.~5 on a
semilogarithmic scale, and the exponential decay is clearly seen.

\begin{figure}[htb]
\vspace*{-1cm}
\centerline{\psfig{width=100mm,angle=90,file=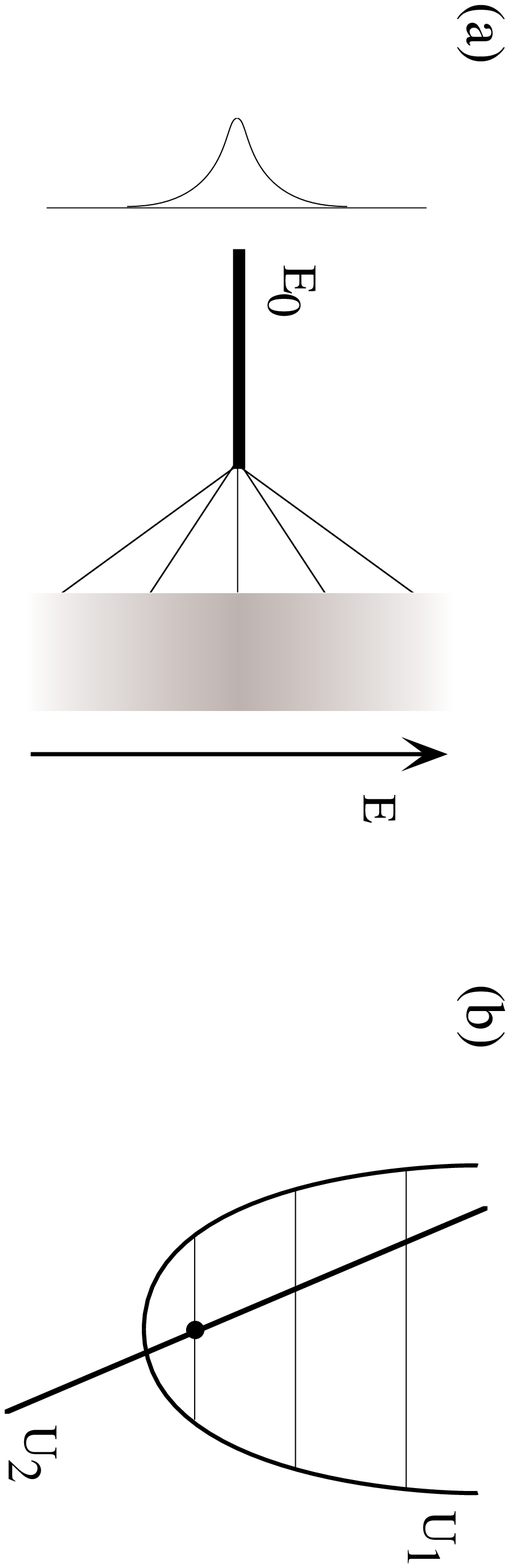}}
\vspace*{-1cm}
\caption{Fig. 3: The basic model for (a) a discrete state coupled to a 
continuum,
and for (b) a molecular representation of the same situation. In (b) the
discrete state is the lowest vibrational state of the harmonic potential, and
the eigenstates of the linear potential form the continuum.
\label{PSS1}}
\end{figure}

\begin{figure}[htb]
\vspace*{-2cm}
\centerline{\psfig{width=80mm,file=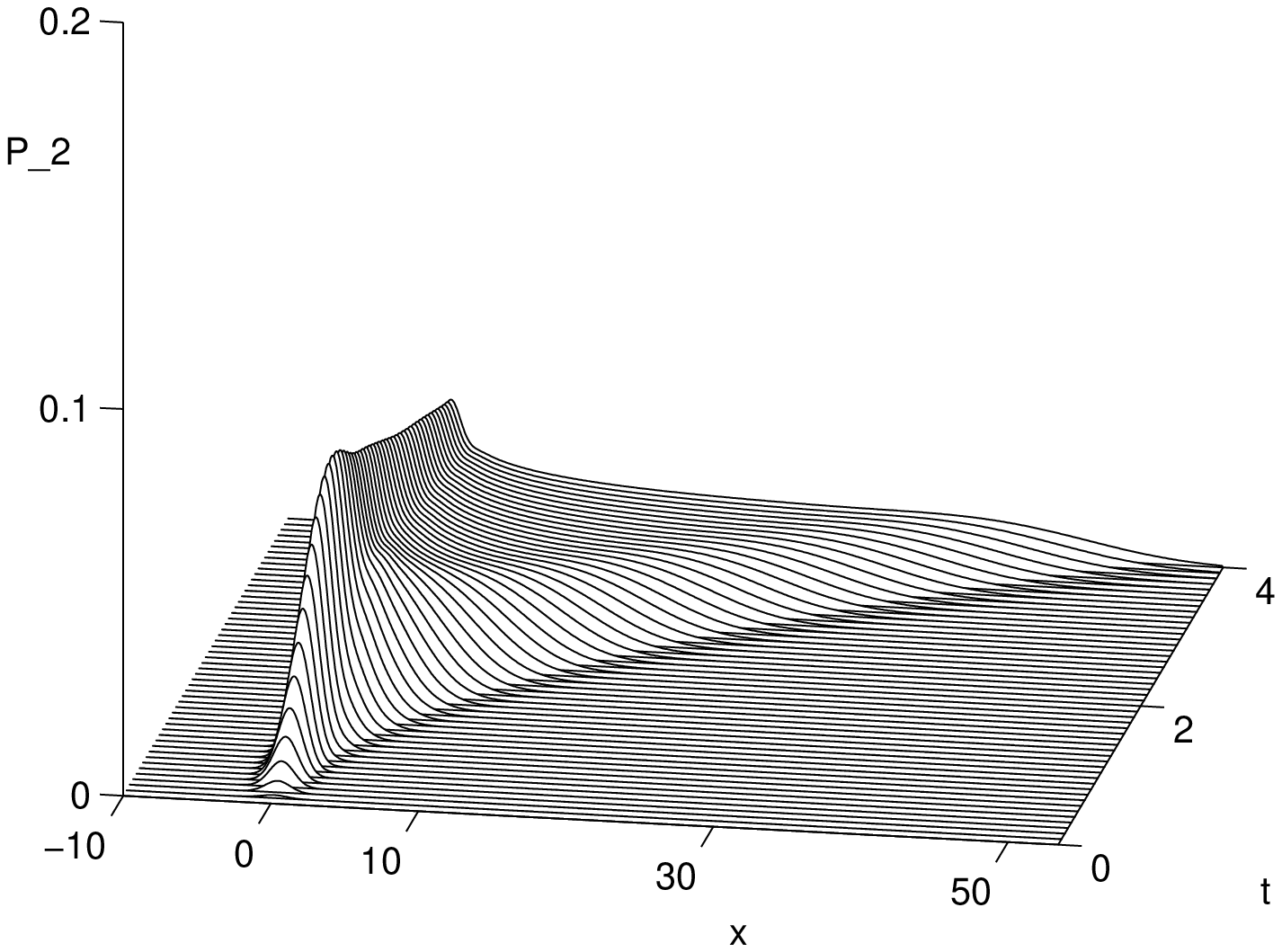}}
\caption{Fig. 4: The excited state wave packet $P_2(x,t)$ for
$\Gamma\simeq 0.26$.\label{SP6}}
\end{figure}

\begin{figure}[htb]
\centerline{\psfig{width=60mm,file=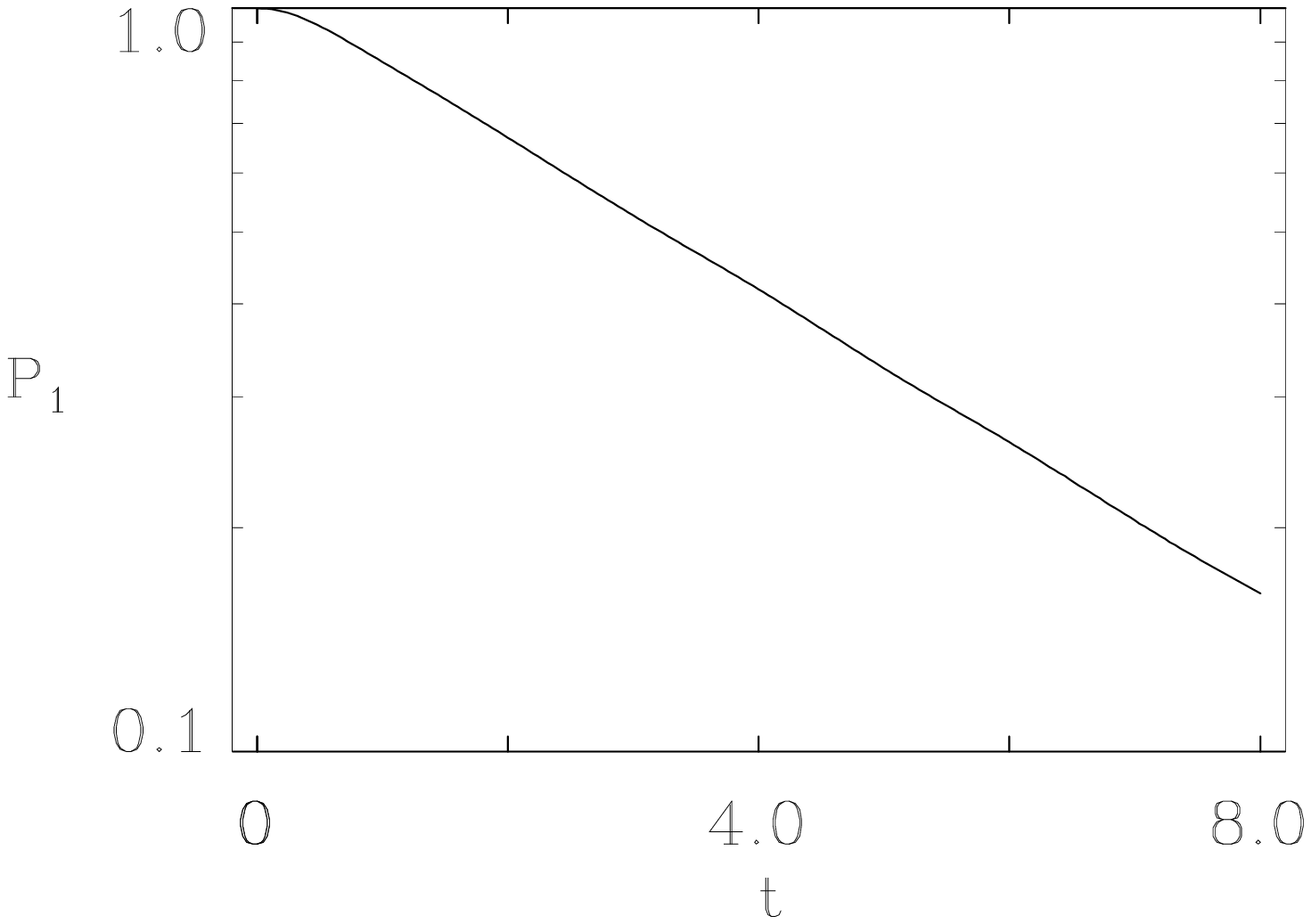}}
\caption{Fig. 5: The ground state population $P_1(t)$ as a function of time
$t$. It shows the exponential nature of the excitation process.\label{SP5}}
\end{figure}

In the model~(\ref{f3}), we can calculate the decay rate in the
Weisskopf-Wigner limit including an analytic expression for the Condon
factor. In the limit of a steep slope, the reflection principle gives
the decay rate
\begin{equation}
   \Gamma =2\pi \frac{V^2}{\alpha \,\left( 2\pi ^2\right) ^{1/4}};  \label{f4}
\end{equation}
this limit of large $\alpha $ is the perturbative regime, and we have found
that the exponential decay can be found for values
\begin{equation}
   \Gamma \leq 1.  \label{f5}
\end{equation}
Figure 4 is computed at $\Gamma \simeq 0.26$, which is well into the
perturbative regime. In our paper~\cite{b10}, we show that the initial wave
packet is found to decay exponentially with a decay rate given by
\begin{equation}
   \Gamma =2\pi V^2| S |^2,  \label{f6}
\end{equation}
where $S$ is the Franck-Condon factor, as long as~(\ref{f5}) is satisfied. For
larger values of $\Gamma$, the decay acquires oscillational features, the
behaviour goes over into the Rabi-type oscillations described above. 

Even when the decay is no longer exponential, the emerging state on the
slope still has a main component shaped like a wave packet, Fig.~6 
where we have $\Gamma \simeq 2.4$, well above the perturbative
regime. Here we can also see the Rabi oscillations between the ground state
and the excited one. This displays the freezing effect discussed above,
which can also be looked upon as wave packet trapping in the adiabatic
potential wells formed by the two intersecting levels; see also~\cite{b2}.
This probability sloshes around in the metastable well and oozes out as a
long tail of probability clearly visible in Fig.~6. For a detailed
discussion of the behaviour, we refer to Ref.~\cite{b10}.

\begin{figure}[htb]
\vspace*{-2cm}
\centerline{\psfig{width=80mm,file=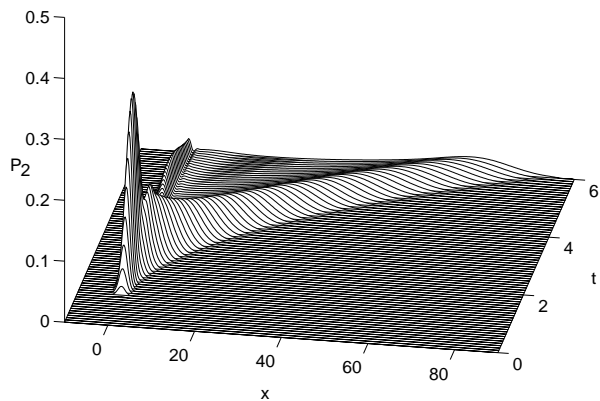}}
\caption{Fig. 6: The excited state wave packet $P_2(x,t)$ for
$\Gamma\simeq 2.4$.\label{PSS8}}
\end{figure}

\section{Conclusions}\label{concl}

We have considered the nuclear wave packet on an electronic energy level in
a molecule as a prototype laboratory for wave packet investigations. The
inital state is well defined, it can be excited and manipulated in a well
controlled mannner, and the time scales are such that the evolution can be
followed in real time. We can combine numerical and analytic investigations
to explore a multitude of quantum effects in the dynamic regime, and compare
the results with well known quantum models. 

We have investigated level crossings, pulsed excitation, metastable
trapping, Rabi flopping, state freezing, wave packet decay, and the effects
of spontaneous decay. In all these cases fundamental concepts of Quantum
Mechanics are involved. This justifies calling the molecular wave packet a
universal quantum object.

Finally a word of warning is needed. If we look at the freezing effect shown
in Fig.~2, we may be tempted to think that the passing 
of the wave packet
back to the ground state, constitutes a measurement if it is still there;
thus the process would be an instance of the Zeno effect in Quantum
Mechanics. However, no recording has been made, no information has been
gathered, and the process is not a measurement. The freezing effect is only a
dynamical consequence of the harmonic ground state to return the wave packet
to its center. By visiting this potential frequently, the state is prevented
from spreading or sliding on the upper level, and the freezing ensues as a
purely dynamical effect.

\end{document}